\documentclass[preprint]{revtex4}

\usepackage{graphicx}
\usepackage{psfrag}

\begin{document}

\title{\bf Measuring atomic NOON-states and using them to make precision measurements}
\author{David W. Hallwood, Adam Stokes, Jessica J. Cooper and Jacob Dunningham}
\affiliation{School of Physics and Astronomy, University of Leeds, Leeds, LS2 9JT, United Kingdom.}

%<<<<<<<<<<<<<<<<<<<<<<<<<<<<<<<<<<<<<<<<<<<<<<<<
\begin{abstract}
A scheme for creating NOON-states of the quasi-momentum of ultra-cold atoms has recently been proposed [New J. Phys. \textbf{8}, 180 (2006)]. This was achieved by trapping the atoms in an optical lattice in a ring configuration and rotating the potential at a rate equal to half a quantum of angular momentum . In this paper we present a scheme for confirming that a NOON-state has indeed been created. This is achieved by spectroscopically mapping out the anti-crossing between the ground and first excited levels by modulating the rate at which the potential is rotated. Finally we show how the NOON-state can be used to make precision measurements of rotation.
\end{abstract}
%<<<<<<<<<<<<<<<<<<<<<<<<<<<<<<<<<<<<<<<<<<<<<<<<
\maketitle

%===============
\section{Introduction}
Mesoscopic superpositions are invaluable for understanding fundamental physics and developing quantum technologies such as precision measurements and quantum computing. Here we study a special type of mesoscopic superposition, the NOON-state~\footnote{The NOON-state is also called the GreenbergerÐHorneÐZeilinger-state (GHZ-state)}, which is a superposition of all particles being in one mode and all particles being in another. We use it to look at two aspects of macroscopic quantum systems: distinguishing a quantum mechanical system from a classical system and making precision measurements.

Recent work in superconductor physics has shown superpositions of flux through a superconducting loop, or superpositions of opposite mesoscopic currents flowing around a loop~\cite{friedman_00, wal_00}. They spectroscopically mapped out an anti-crossing between the ground and first excited states of the system by resonantly coupling them with an oscillating magnetic field. The anti-crossing gives clear evidence of entanglement in the system. Here we look at simulating the superconducting system with neutral ultra-cold atoms trapped in a ring trapping potential created with a light field. The effect of the magnetic field is reproduced for neutral atoms by rotating the trapping potential. 

Modulating optical potentials has already produced several significant results in ultra-cold atom systems. St\"{o}ferle \emph{et al.} demonstrated the superfluid to mott Insulator transition using this technique~\cite{stoferle_04, greiner_02}. Also, Volz \emph{et al.} demonstrated the creation of molecules in an optical lattice~\cite{volz_06}. Transitions between optical lattice sites are induced by modulating the intensity of the optical lattice at the correct frequency. When molecules are formed only the tunneling of pairs of atoms is observed. While that work considered the tunneling of atoms between sites, here we modulate the rotation rate of the lattice which allows us to study the transition of atoms between different quasi-momentum modes. This is of particular importance at the moment with the growing experimental interest in rotating systems~\cite{ryu_07, henderson_09}.

The large amount of quantum entanglement in NOON-states means that they should be able to make measurements that go beyond the shot noise limit of $1/\sqrt{N}$, and potentially achieve the Heisenberg limit of $1/N$~\cite{wineland_92, wineland_94, bollinger_96}, where $N$ is the number of particles. In this paper we shall consider a scheme for making precision measurements of rotation.

The paper is split into four parts. Initially we will review how ultra-cold atoms can be used to make a NOON-state of superfluid flow. We will then show how NOON-states can be demonstrated by resonantly coupling the ground and first excited state by modulating the rotation rate at specific frequencies. Finally we will look at measuring an applied rotation and how the energy gap can be measured when a NOON-state is formed. An understanding of the size of the energy gap is of particular importance because it tells us how well the NOON-state is protected from excitations that will destroy it~\cite{dagnino_09}. 

%===============
\section{The system}

%Figure system
%>>>>>>>>>>>>>>>>>>>>>>>>>>>>>>>>>>>>>>>>>>>>>>>>>>>>>>
\begin{figure}
\includegraphics[width=7cm]{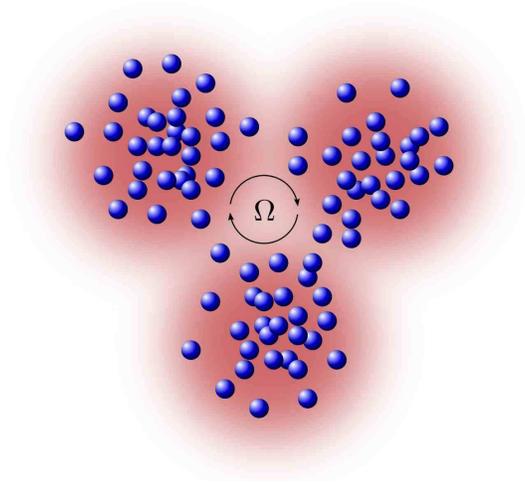}
\caption{\small Schematic of the system. Atoms are trapped by the optical dipole force from an optical lattice loop of three sites. The lattice sites are rotated at a rate which we describe by the rotation phase $\Omega$. The atoms interact on the sites and can tunnel between sites. When $\Omega=\pi$ and the tunneling is large a NOON-state is created. By modulating the rotation phase it is possible to show a NOON-state has been created.}
\label{fig:system}
\end{figure}
%<<<<<<<<<<<<<<<<<<<<<<<<<<<<<<<<<<<<<<<<<<<<<<<<<<<<<<

The schematic of the system we consider is shown in figure~\ref{fig:system}. Ultra-cold atoms are trapped by the optical dipole force in an optical lattice loop of three sites. The sites are rotated at a rate $v=(\hbar/m)\Omega/L$, where $L$ is the circumference of the loop and $\Omega$ is the applied phase around the loop, which we call the rotation phase, and the quantity that will be used in this paper to describe the rotation~\cite{rey_03}. There have already been several experiments that have demonstrated these types of potentials~\cite{henderson_09, houston_08}.

We will assume atoms only interact when they are on the same site and with strength $U$, and the atoms can also tunnel to adjacent sites. One of the barriers is taken to be higher than the other two and is parameterized by the coupling rate $J-\Delta J$. The lower two barriers each have tunneling rate $J$. As described in~\cite{hallwood_06}, the different barrier heights break the symmetry of the system and allow NOON-states to be created for all numbers of atoms. This system can be described by the Bose-Hubbard model~\cite{jaksch_98}. In terms of the quasi-momentum basis the Hamiltonian is~\cite{hallwood_07},
\begin{eqnarray}
\hat{H}_0 &=& \hat{H}_U + \hat{H}_J + \hat{H}_{\Delta J} \nonumber\\
\hat{H}_U &=& \frac{U}{3} \left[\hat{n}_{0}(\hat{n}_{0}-1) +  \hat{n}_{1}(\hat{n}_{1}-1) + \hat{n}_{-1}(\hat{n}_{-1}-1)\right.  \nonumber\\
&&+ 4\left(\hat{n}_{0}\hat{n}_{1} + \hat{n}_{0}\hat{n}_{-1} + \hat{n}_{1}\hat{n}_{-1} \right) +\left. 2\left(\hat{\alpha}^2_{0}\hat{\alpha}_{1}^{\dag}\hat{\alpha}_{-1}^{\dag} + \hat{\alpha}^2_{1}\hat{\alpha}_{0}^{\dag}\hat{\alpha}_{-1}^{\dag} + \hat{\alpha}^2_{-1}\hat{\alpha}_{1}^{\dag}\hat{\alpha}_{0}^{\dag} + \mbox{h.c.} \right) \right]\nonumber\\
 \hat{H}_J&=&-2J \left( \hat{n}_{0} \cos\left(\frac{\Omega}{3}\right) + \hat{n}_{1} \cos\left(\frac{\Omega-2\pi}{3}\right) + \hat{n}_{-1} \cos\left(\frac{\Omega+2\pi}{3}\right)\right) \nonumber \\
 \hat{H}_{\Delta J}&=&\frac{2\Delta J}{3}\left( \hat{n}_{0} \cos\left(\frac{\Omega}{3}\right) + \hat{n}_{1} \cos\left(\frac{\Omega-2\pi}{3}\right) + \hat{n}_{-1} \cos\left(\frac{\Omega+2\pi}{3}\right)\right. \nonumber\\
&& \left. + (\hat{\alpha}_{0}^{\dag}\hat{\alpha}_{1}+\hat{\alpha}_{1}^{\dag}\hat{\alpha}_{0})\cos\left(\frac{\Omega+2\pi}{3}\right)+(\hat{\alpha}_{1}^{\dag}\hat{\alpha}_{-1}+\hat{\alpha}_{-1}^{\dag}\hat{\alpha}_{1})\cos\left(\frac{\Omega}{3}\right)\nonumber \right.\\
&&\left.+(\hat{\alpha}_{-1}^{\dag}\hat{\alpha}_{0} +\hat{\alpha}_{0}^{\dag}\hat{\alpha}_{-1})\cos\left(\frac{\Omega-2\pi}{3}\right)\right)
\label{eq:H0}
\end{eqnarray}
where $\hat{\alpha}_{k}^{\dag}$ ($\hat{\alpha}_{k}$) creates (destroys) an atom in momentum mode $k$ and $\hat{n}_{k}$ is the operator for the number of atoms in momentum mode $k$. The position of the atoms are limited to three sites, so the quasi-momentum (or flow) of the atoms is limited to three modes: not moving, $k=0$, moving with one quantum of clockwise flow, $k=1$, and moving with one quantum of anti-clockwise flow $k=-1$, where a quantum of flow is $v_0=(\hbar/m)(2\pi/L)$ or a $2\pi$ phase around the loop.

%Figure showing the NOON state
%>>>>>>>>>>>>>>>>>>>>>>>>>>>>>>>>>>>>>>>>>>>>>>>>>>>>>>
\begin{figure}
\includegraphics[width=7cm]{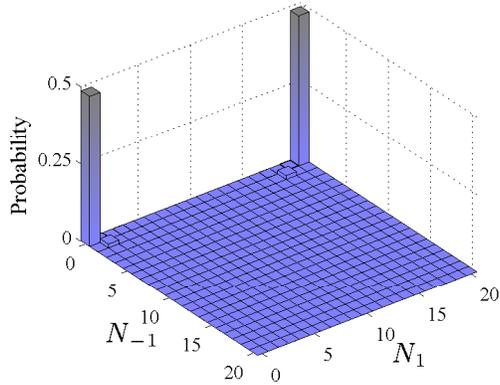}
\caption{\small Plot of the ground state of a system of 30 atoms in a NOON-state created by applying a rotation phase $\Omega=\pi$, $U/J=0.05$ and $\Delta J/J=0.01$. Each bar represents the probability of finding the ground state in state $|N_{-1}, N-N_{1}-N_{-1}, N_{1}\rangle$, where $N$ is the total number of atoms in the system.}
\label{fig:SP}
\end{figure}
%<<<<<<<<<<<<<<<<<<<<<<<<<<<<<<<<<<<<<<<<<<<<<<<<<<<<<<

It has previously been shown that a NOON-state of quasi-momentum can be produced by precisely controlling the system~\cite{hallwood_06, dunningham_06, hallwood_07, nunnenkamp_08}. Here this is done by applying a rotation phase $\Omega=\pi$ to the system, where the rate of tunneling is large enough to allow the atoms to tunnel freely between the sites and the interactions are strong enough to lift the degeneracy at $\Omega=\pi$. For this rotation phase, the 0 and $1$ quasi-momentum modes are degenerate, so the ground and first excited states are,
\begin{eqnarray}
|+\rangle &=& \frac{1}{\sqrt{2}}\left(|0,N,0\rangle + |0,0,N\rangle \right), \nonumber \\
|-\rangle &=& \frac{1}{\sqrt{2}}\left(|0,N,0\rangle - |0,0,N\rangle \right),
\end{eqnarray}
which are both NOON-states. Here we describe the state of the system in the occupation basis where each term in the ket represents the number of atoms in the -1, 0 and 1 quasi-momentum modes. By changing the parameters in the Hamiltonian given in equation~\ref{eq:H0} we can produce a state $|\Psi\rangle$ that is close to $|+\rangle$, in the sense that its fidelity with $|+\rangle$ is close to 1, $|\langle +|\Psi \rangle|\approx 1$. An example of a NOON-state for a system of 30 atoms is shown in figure~\ref{fig:SP}. The two bars show there is an equal probability of finding the system in states $|0,N,0\rangle$ and $|0,0,N\rangle$ with small probabilities of finding the system in other states. We must now develop an experimentally accessible scheme that can distinguish this state from a classical mixture.

%========================================
\section{Demonstration of a NOON-state}
\label{sec:NOON}
%Anti-crossing
%>>>>>>>>>>>>>>>>>>>>>>>>>>>>>>>>>>>>>>>>>>>>>>>>>>>>>>
\begin{figure}
\includegraphics[width=7cm]{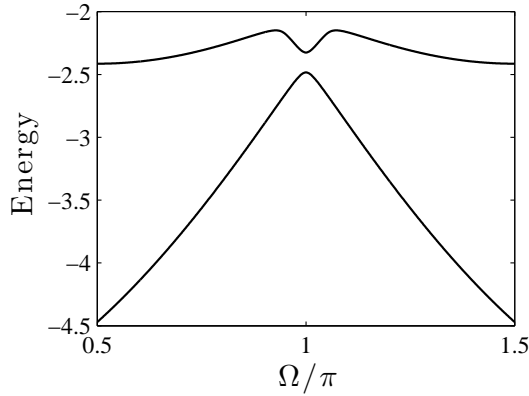}
\caption{\small Energy of the ground and first excited state as a function of rotation phase for a system of 3 atoms. We see a clear anti-crossing at $\Omega=\pi$ suggesting a NOON-state has been created.}
\label{fig:EvOmega}
\end{figure}
%<<<<<<<<<<<<<<<<<<<<<<<<<<<<<<<<<<<<<<<<<<<<<<<<<<<<<<

%What is an anti-crossing
Here we will demonstrate the NOON-state by mapping out an anti-crossing between the ground and first excited states as we increase the rotation phase applied to the system. The anti-crossing provides clear evidence of the superposition. For example, consider a two state system, $|0\rangle$ and $|1\rangle$, where the energy of these states are given by $\varepsilon_0=x$ and $\varepsilon_1=-x$, and $x$ is some external parameter, such as rotation phase. The system is described by the Hamiltonian $H=\varepsilon_0 |0\rangle \langle 0|+\varepsilon_1 |1\rangle \langle 1 | +\Delta (|0\rangle \langle 1 | + |1\rangle \langle 0 |)$, where $\Delta$ is the coupling strength between the two states. When $x=0$ the two states are degenerate if $\Delta=0$, but the degeneracy is lifted if $\Delta$ is non-zero giving the two eigenstates $\frac{1}{\sqrt{2}}\left( |0\rangle \pm |1\rangle \right)$ with an energy difference $2\Delta$. 

%Single or mac superposition
The same idea applies to our system where the rotation changes the energy of the different eigenstates. The coupling between $|0,N,0\rangle$ and $|0,0,N\rangle$ lifts the degeneracy when $\Omega=\pi$ creating an anti-crossing (see figure~\ref{fig:EvOmega}). The ground state of the system is approximately,
\begin{eqnarray}
|\Psi_M\rangle &=& \frac{1}{\sqrt{2 N!}}\left( \left(\hat{\alpha}_{0}^{\dag}\right)^N + \left(\hat{\alpha}_{1}^{\dag}\right)^N\right)|\mbox{vac}\rangle  \nonumber \\
&=& \frac{1}{\sqrt{2}} \left( |0,N,0\rangle + |0,0,N\rangle \right)
\label{eq:NOON}
\end{eqnarray}
but we must be careful that we do not get single particle superpositions,
\begin{eqnarray}
|\Psi_{SP}\rangle &=& \frac{1}{2^{N/2} \sqrt{N!}}\left( \hat{\alpha}_{0}^{\dag} +\hat{\alpha}_{1}^{\dag}\right)^N|\mbox{vac}\rangle \nonumber \\
 &=& \frac{1}{2^{N/2}}\sum_{m=0}^N \sqrt{\frac{N!}{m!(N-m)!}} |0,m,N-m\rangle
\end{eqnarray}
The measurement outcomes of these two states are very different, so they should be easy to distinguish.

We will now show how the energy gap can be measured spectroscopically by modulating the lattice at specific frequencies. If the modulating frequency, $\omega$, is on resonance, so $\Delta E = \hbar \omega$, then the ground and excited states become coupled. By scanning across the applied rotation phase we can map out the anti-crossing by finding the resonant frequencies. Furthermore, if we find there is an energy gap at rotation phase $\Omega=\pi$ and we only measure the system in the states $|0,N,0\rangle$ and $|0,0,N\rangle$ with equal probability then we know we have created a NOON-state, because a classical mixture of $|0,N,0\rangle$ and $|0,0,N\rangle$ produces no gap.

%-----------------------------------------
\subsection{Modeling the system}
%The interaction picture
We remove the time dependence of the modulation by moving into the interaction picture. Here we simplify the system by considering only the lowest three eigenstates, for example, in our system with $\Omega=\pi$ the lowest three eigenstates are approximately $|0\rangle=|+\rangle$, $|1\rangle=|-\rangle$ and $|2\rangle=(|0,N-1,1\rangle+|0,1,N-1\rangle)/\sqrt{2}$. The effective Hamiltonian is,
\begin{eqnarray}
\hat{H}_{E}&=&\sum_{k=0}^{3} \varepsilon_{k} |k\rangle\langle k| + \sum_{k_1\ne k_2} V_{k_1k_2} |k_1\rangle\langle k_2| \cos(\omega t) \nonumber \\
          &=& \hat{H}_{0} + \hat{V}\cos(\omega t)
\end{eqnarray}
where $\hat{H}_0=\sum_{k=0}^{3} \varepsilon_{k} |k\rangle\langle k|$ is the Hamiltonian of the system before the extra modulation is turned on with rotation phase $\Omega$, $\hat{V}$ is the coupling term created by the modulations and $\omega$ is the modulation frequency. In the interaction picture the Hamiltonian becomes,
\begin{eqnarray}
\hat{H}_{I} &\approx&  -\Delta_{01} |1\rangle\langle 1| - \Delta_{02} |2\rangle\langle 2| \nonumber\\
                       &&+ \frac{V_{01}}{2}(|0\rangle\langle 1|+|1\rangle\langle 0|) + \frac{V_{02}}{2}(|0\rangle\langle 2|+|2\rangle\langle 0|) + \frac{V_{21}}{2}(|2\rangle\langle 1|+|1\rangle\langle 2|)
\end{eqnarray}
Now the system only depends on the coupling, $V_{k_1 k_2}$, between the eigenstates of $\hat{H}_0$ and the detuning of the modulation frequency, $\Delta_{k_1 k_2}$. To make a good measurement of the energy difference, $\Delta E$, between states $|0\rangle$ and $|1\rangle$ we want near resonant coupling between them, $\Delta_{01} \approx 0$, and far off resonant coupling between $|0\rangle$ and $|2\rangle$, $\Delta_{02} \gg \Delta_{01}$. We also require $V_{01}\ne 0$ and preferably to be large and a sharp resonance from the coupling.

%Our system in the interaction picture
When we apply the modulations to our system the coupling term is, 
\begin{eqnarray}
\hat{V} \cos(\omega t)&=& \left\{ \left(\frac{2J}{3}-\frac{2\Delta J}{9}\right)\left(\hat{n}_{0}\sin\left(\frac{\Omega}{3}\right)+\hat{n}_{1}\sin\left(\frac{\Omega-2\pi}{3}\right)+\hat{n}_{-1}\sin\left(\frac{\Omega+2\pi}{3}\right)\right)\right. \nonumber\\
&&-\frac{2\Delta J}{9}\left((\hat{\alpha}_{0}^{\dag}\hat{\alpha}_{1}+\hat{\alpha}_{1}^{\dag}\hat{\alpha}_{0})\sin\left(\frac{\Omega_0+2\pi}{3}\right)+(\hat{\alpha}_{1}^{\dag}\hat{\alpha}_{-1}+\hat{\alpha}_{-1}^{\dag}\hat{\alpha}_{1})\sin\left(\frac{\Omega_0}{3}\right)\nonumber \right.\\
&&\left. \left.+(\hat{\alpha}_{-1}^{\dag}\hat{\alpha}_{0} +\hat{\alpha}_{0}^{\dag}\hat{\alpha}_{-1})\sin\left(\frac{\omega_0-2\pi}{3}\right)\right)\right\} A\cos(\omega t)
\label{eq:HI}
\end{eqnarray}
where we have assumed $\cos(A\cos(\omega t)/3)\approx 1$ and $\sin(A\cos(\omega t)/3)\approx A\cos(\omega t)/3$.

%Calculating the coupling
At $\Omega=\pi$, and assuming the ground state has a fidelity with $|+\rangle$ that is close to 1, $\langle 1 | \hat{V} | 0\rangle = V_{01} = \left( 2J - \frac{2\Delta J}{3} \right)\frac{N\Delta\omega}{\sqrt{3}}$ and $\langle 2 | \hat{V} | 0\rangle = \langle 2 | \hat{V} | 1\rangle = V_{12} = 0$. This suggests it is very easy to measure the energy difference in this case. When $|0\rangle\approx |0,N,0\rangle$ and $|1\rangle\approx |0,N-1,1\rangle$ or $|0\rangle\approx |0,0,N\rangle$ and $|1\rangle\approx |0,1,N-1\rangle$ the coupling is $\langle 1 | \hat{V} | 0\rangle = V_{01} = - \frac{2\Delta JN\Delta\omega}{9} \sin\left( \frac{\omega_{0}+2\pi}{3} \right)$. This is much smaller than the case when we have a NOON-state and it will be harder to distinguish the two states because they only differ by a single atom. We will show how to get around this problem below. For the states that are between these two limits we must calculate the coupling numerically as we will show.

%-----------------------------------
\subsection{Measurement procedure}
%Experimentally making the system
%>>>>>>>>>>>>>>>>>>>>>>>>>>>>>>>>>>>>>>>>>>>>>>>>>>>>>>
\begin{figure}
\includegraphics[width=7cm]{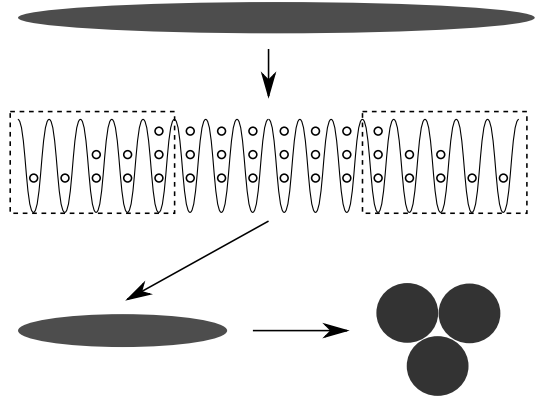}
\caption{\small To perform precision measurements we need to know the number of atoms in the system. This can be done by first adiabatically moving into the Mott state. The atoms at the centre of the trap with a fixed number are kept while the edge atoms are removed. The number of remaining atoms can be chosen in this way. The potential is then adiabatically evolved into the desired shape.}
\label{fig:scheme}
\end{figure}
%<<<<<<<<<<<<<<<<<<<<<<<<<<<<<<<<<<<<<<<<<<<<<<<<<<<<<<

Many measurements must be made on an ensemble of systems to determine whether there is an anti-crossing and the number of atoms in each system must be the same. Figure~\ref{fig:scheme} shows one possible way this could be achieved. The atoms are initially confined to a cigar shaped trap. An optical lattice is adiabatically applied forcing a definite number of atoms in each of the sites. Volz \emph{et al.} has already experimentally demonstrate how exactly two atoms can be put on each lattice site~\cite{volz_06}. By removing the atoms from the edge sites we fix the total number of atoms. The lattice is then removed and the trapping potential deformed to the three site configuration we want.

\begin{figure}
\includegraphics[width=7cm]{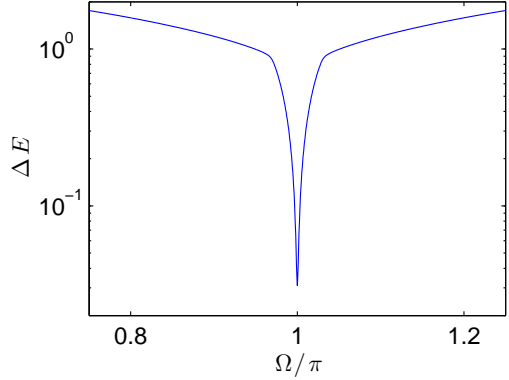}
\caption{\small Energy difference between the ground and first excited state, $\Delta E$ as a function of rotation phase. We can see the gap never goes to zero, suggesting a NOON-state is formed at $\Omega=\pi$.}
\label{fig:DEvOmega}
\end{figure}

We now want to measure the energy difference between the ground and first excited state for different applied rotation phases, $\Omega$. This is done by modulating the rotation phase at the frequency, $\omega$, about $\Omega$, which couples the ground and excited states. The procedure is as follows: (1) pick an initial rotation phase $\Omega$ and apply a modulation frequency $\omega$ for a time $t$. (2) Now adiabatically change $\Omega$ to a region where the ground and excited states are known to be $|0,N,0\rangle$ and $|0,0,N\rangle$. This occurs when we are close to the anti-crossing but not exactly on it so the superposition has not yet formed. (3) Measure the state of the system, and, (4) repeat for different times, $t$, until we can see whether the system is resonantly oscillating between the ground and excited state. If we have chosen the resonant frequency we know the energy gap at this rotation phase. If not we repeat for different modulation frequencies until we do. (5) Next we repeat for different rotation phases until we get a plot like~\ref{fig:DEvOmega}. The sharp drop in $\Delta E$ is where the ground and excited states only involve superpositions of $|0,N,0\rangle$ and $|0,0,N\rangle$. If the energy gap is not zero at $\Omega=\pi$ then we know the system can be used to create a NOON-state.

%Measuring the energy gap
 %>>>>>>>>>>>>>>>>>>>>>>>>>>>>>>>>>>>>>>>>>>>>>>>>>>>>>>
%<<<<<<<<<<<<<<<<<<<<<<<<<<<<<<<<<<<<<<<<<<<<<<<<<<<<<<

%Resonance width and coupling strength
%>>>>>>>>>>>>>>>>>>>>>>>>>>>>>>>>>>>>>>>>>>>>>>>>>>>>>>
\begin{figure}
\includegraphics[width=8cm]{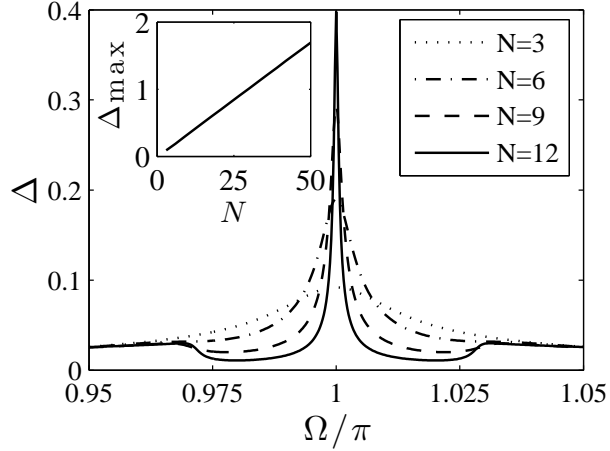}
\caption{\small Plot shows shows how the coupling strength changes as a function of rotation phase for different numbers of atoms. We can see there is an increase in the peak as the number of atoms increases. The subplot shows the change in the maximum coupling strength as a function of the number of atoms for $\Omega=\pi$.}
\label{fig:coupling}
\end{figure}
%<<<<<<<<<<<<<<<<<<<<<<<<<<<<<<<<<<<<<<<<<<<<<<<<<<<<<<

The rate at which the amplitudes oscillate between the ground and excited states (Rabi frequency) is proportional to the coupling strength. Previously we saw that the coupling strength increases when the ground state is a NOON-state, so the Rabi frequency will also increase. This gives a sharp signature showing when a NOON-state has been created. Figure~\ref{fig:coupling} shows how the strength of the coupling (and therefore the Rabi frequency) increases when we have a NOON-state and the resonance becomes sharper as we increase the number of atoms. There is also a dip in the Rabi frequency either side of the peak. This is the region when the ground and excited states are approximately $|0,N,0\rangle$ and $|0,0,N\rangle$, which are very weakly coupled.

In summary, we can measure whether we have a NOON-state by finding the modulating frequency that couples the ground and excited state at that rotation phase. This will then show whether there is an anti-crossing, which is a clear signature of a superposition. We also find at the NOON-state the Rabi frequency rapidly increases, which is a clear signature that we have created a NOON-state.

%Making a precisely rotating system
A direct application of this system is to control the rotation of an object. The scheme gives a reference to relate all other rotations. We would start with a small number of atoms to get an approximate initial rotation rate and then improve the accuracy by repeating the scheme with more atoms.

%\subsection{Experimental considerations}

%=================================================
\section{Precision measurement using NOON-states}
The quantum properties of a NOON-state allow precision measurements far greater than classical measurements. Schemes have already been proposed for using atomic NOON-states in sites to make precision measurements~\cite{dunningham_01, dunningham_04}, but so far no one has looked at how NOON-states of quasi-momentum can be used. Here we show they can be used to make precision measurements of rotation. 

We want to measure an increase or decrease in the rate of rotation, which is equivalent to an additional rotation phase. We start with a NOON-state created as described above, then non-adiabatically apply the rotation phase we want to measure, $\Delta \Omega$, and let the system evolve for a time $t$.  The wavefunction is no longer an eigenstate of the Hamiltonian with the additional rotation phase, $\hat{H}'$, so the wavefunction becomes,
\begin{eqnarray}
|\Psi(t)\rangle =\frac{1}{\sqrt{2}}\left( |0,N,0\rangle+e^{-i\Delta E t}|0,0,N\rangle \right)
\label{eq:WF1}
\end{eqnarray}
where $\Delta E = E_{00N}-E_{0N0}$, $\hat{H}' |p,n,m\rangle = E_{pnm}|p,n,m\rangle$ and we assume $H'$ does not strongly couple $|0,N,0\rangle$ and $|0,0,N\rangle$ to other states.

The additional rotation, $\Delta \Omega$, is then removed, leaving a phase difference between the states $|0,N,0\rangle$ and $|0,0,N\rangle$. To read out the additional rotation phase we must convert the phase difference into an amplitude. This is done by adiabatically changing the rotation phase to a value where the ground and excited states are known to be $|0,N,0\rangle$ and $|0,0,N\rangle$, giving,
\begin{eqnarray}
|\Psi(t)\rangle = \cos\left(\frac{\Delta E}{2} t \right) |0,N,0\rangle + i \sin\left(\frac{\Delta E}{2} t \right) |0,0,N\rangle
\end{eqnarray}
The additional rotation phase can now be measured by looking at the probability of finding the system in state $|0,N,0\rangle$ and $|0,0,N\rangle$ and using,
\begin{eqnarray}
\Delta E = \sqrt{3}N\left(J-\frac{\Delta J}{3}\right)\sin\left( \Delta \Omega\right/3)
\end{eqnarray}
Note that $\Delta E$ is proportional to $N$ so the uncertainty has reached the Heisenberg limit of $1/N$, rather than the shot noise limit of $1/\sqrt{N}$~\cite{wineland_92, wineland_94, bollinger_96}.

We should note that the additional rotation phase is only applied for a time $t$ and not when we are creating the NOON-state and adiabatically evolving at the end. This can be achieved by initially creating the NOON-state with the systems rotation axis perpendicular to the rotation axis we want to measure, then non-adiabatically evolving the system so the systems rotation axis is parallel to the rotation axis we want to measure.

%==========================================================
\section{Measuring the energy gap at $\Omega=\pi$}
In section~\ref{sec:NOON} we showed how to measure the energy gap for different values of $\Omega$, but it required several experimental steps. Here we describe a scheme for measuring the energy gap at $\Omega=\pi$, which will be experimentally simpler although it can only measure the gap at one value of $\Omega$. Nunnenkamp \emph{et al.} described a method for measuring the energy gap by starting at $\Omega=0$ then rapidly changing the rotation phase to $\Omega=\pi$~\cite{nunnenkamp_08}. The state of the system then oscillates between $|0,N,0\rangle$ and $|0,0,N\rangle$ with a frequency equal to the energy gap. However, the large change in $\Omega$ means there is strong coupling to higher energy levels, so gives poor resolution and the large change in rotation would be experimentally difficult to achieve without excitations to higher energy levels. Here we show how a smaller change in $\Omega$ can achieve higher resolution.

Again we create a NOON-state by applying a rotation of $\Omega=\pi$, $|\Psi(t=0)\rangle = (|0,N,0\rangle + |0,0,N\rangle)/\sqrt{2}$. As before an additional rotation, $\Omega'$, is non-adiabatically applied, which does not have to be known. Under the new Hamiltonian the system evolves as described by equation~\ref{eq:WF1}.

%Probability curve
%>>>>>>>>>>>>>>>>>>>>>>>>>>>>>>>>>>>>>>>>>>>>>>>>>>>>>>
\begin{figure}
\includegraphics[width=8cm]{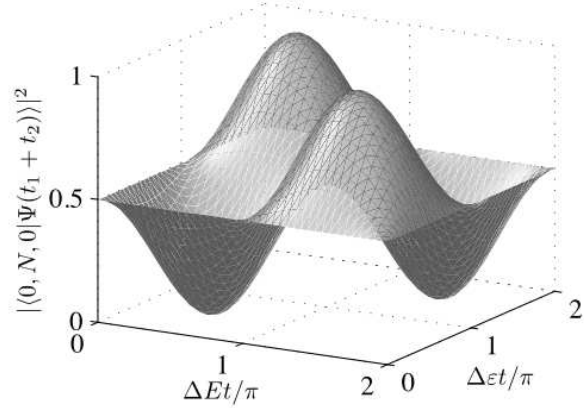}
\psfrag{Probability}{Probability}
\caption{\small Probability of finding all the atoms in state $|0,N,0\rangle$ as a function of two times the system is held for.}
\label{fig:DEandDep}
\end{figure}
%<<<<<<<<<<<<<<<<<<<<<<<<<<<<<<<<<<<<<<<<<<<<<<<<<<<<<<

Again after a set time, $t_1$, the rotation is non-adiabatically removed. This time we do not adiabatically rotate to a value of $\Omega$ where the ground and excited states are $|0,N,0\rangle$ and $|0,0,N\rangle$, but allow the system to evolve for a time $t_2$ before measuring the state of the system. The wavefunction evolves to,
\begin{eqnarray}
|\Psi(t_1+t_2)\rangle=\cos\left(\frac{\Delta E}{2} t_1 \right)e^{i\Delta\epsilon t_2} |+\rangle - i \sin\left(\frac{\Delta E}{2} t_1 \right) e^{-i\Delta\epsilon t_2}|-\rangle
\end{eqnarray}
The probabilities of finding the system in the states $|0,N,0\rangle$ and $|0,0,N\rangle$ are,
\begin{eqnarray}
|\langle N,0|\Psi(t_1+t_2)\rangle|^2 = \frac{1}{2} \left( 1-\sin(\Delta E t_1)\sin(\Delta\epsilon t_2)) \right), \nonumber \\
|\langle 0,N|\Psi(t_1+t_2)\rangle|^2 = \frac{1}{2} \left( 1+\sin(\Delta E t_1)\sin(\Delta\epsilon t_2)) \right), \nonumber \\
\end{eqnarray}
By evolving the system for different $t_1$ and $t_2$ we can create figure~\ref{fig:DEandDep} and use this to measure the energy gap at the anti-crossing and the applied rotation $\Omega'$. 

\section{Conclusion}
We have demonstrated how modulating the rate of rotation of an optical lattice loop provides a method for probing the momentum distribution of trapped ultra-cold atoms. Here we use this technique to detect the creation of a special type of mesoscopic superposition, the NOON-state, by mapping out an anti-crossing of the energy gap between the ground and first excited state as a function of the rate of rotation of the trapping potential. We also demonstrate how this system can be used to make precision measurements of rotation using the NOON-state and provide another technique for measuring the energy gap specifically where the NOON-state is created.

%~~~~~~~~~~~~~~~~~~~~~~~~~~~~~~~~~~~~~~~~~~~~~~~~~~~~~~~~~~~~~~~~~~~

%~~~~~~~~~~~~~~~~~~~~~~~~~~~~~~~~~~~~~~~~~~~~~~~~~~~~~~~~~~~~~~~~~~~~

\end{document}